**Article Title: The effects of ENSO, climate change and human activities on the water level of Lake Toba, Indonesia: a critical literature review**

**Article Type: Review**

[1,2] **Hendri Irwandi**, [1] **Mohammad Syamsu Rosid**, [1] **Terry Mart** *

[1] Department of Physics, Faculty of Math and Science, Universitas Indonesia, Kampus Depok City,16424, Indonesia

[2] Deli Serdang Climatology Station, Meteorology Climatology and Geophysical Agency (BMKG), Medan City, North Sumatra, Indonesia

- Email: terry.mart@sci.ui.ac.id

**Abstract**

This research quantitatively and qualitatively analyzes the factors responsible for the water level variations in Lake Toba, North Sumatra Province, Indonesia. According to several studies carried out from 1993 to 2020, changes in the water level were associated with climate variability, climate change, and human activities. Furthermore, these studies stated that reduced rainfall during the rainy season due to the El Niño Southern Oscillation (ENSO) and the continuous increase in the maximum and average temperatures were some of the effects of climate change in the Lake Toba catchment area. Additionally, human interventions such as industrial activities, population growth, and damage to the surrounding environment of the Lake Toba watershed had significant impacts in terms of decreasing the water level. However, these studies were unable to determine the factor that had the most significant effect, although studies on other lakes worldwide have shown these factors are the main causes of fluctuations or decreases in water levels. A simulation study of Lake Toba's water balance showed the possibility of having a water surplus until the mid-twenty-first century. The input discharge was predicted to be greater than the output; therefore, Lake Toba could be optimized without affecting the future water level. However, the climate projections depicted a different situation, with scenarios predicting the possibility of extreme climate anomalies, demonstrating drier climatic conditions in the future. This review concludes that it is necessary to conduct an in-depth, comprehensive, and systematic study to identify the most dominant factor among the three that is causing the decrease in the Lake Toba water level and to describe the future projected water level.

## 1. INTRODUCTION

The word lake comes from the Latin word *lacus*, which means a medium to large "hole" or "space" that is filled with water and surrounded by land (Day and Garratt 2006; Hutter et al. 2010). Lakes are important components of the hydrological cycle that benefit ecosystems and have economic value for human activities (Wantzen et al. 2008; Yao et al. 2018; Davraz et al. 2019; Tao et al. 2020). According to Scott and Huff (1996), Samuelsson et al. (2010), Notaro et al. (2013), and Dehghanipour et al. (2020a, b), lakes have a significant impact on several aspects of the Earth, including the local climate of a region, by supporting ecosystems with hundreds of species of diverse flora and fauna (Awange et al. 2013; Schulz et al. 2020). Lakes also have socioeconomic roles as sources of fresh water, renewable energy, and agricultural irrigation (Guo et al. 2015; Wang et al. 2017; Boadi and Owusu 2019; Pham-Duc et al. 2020; Farrokhzadeh et al. 2020). Therefore, it is important to carry out studies that examine all aspects affecting lakes to support the use of sustainable water resources.

Lake Toba, which formed approximately 75,000 years ago, is the largest tectonic and volcanic lake in the world (Williams 2012; Chesner 2012; Wagner et al. 2013). This lake is geographically located in the middle of the northern part of Sumatra Island, North Sumatra Province, Indonesia. It stretches between 2°21'32"-2°56'28" N and 98°26'35"-99°15'40" E, with water surface and catchment areas of approximately 1124 km$^2$ and 2486 km$^2$, respectively. According to Lukman and Ridwansyah (2011), the lake surface is 903 m above sea level, approximately 50 km in length and 27 km in width, and it has an average depth of 228 m. The Lake Toba area (see Fig. 1) has been recognized as being beneficial for tourism, agriculture, plantation, fisheries, and industry (Saragih and Sunito 2001; Nasution and Damanik 2009), with its discharge used by power plants to supply electricity in North Sumatra (Loebis 1999; Sianturi 2004; Sihotang et al. 2012; Lukman 2017).

Studies carried out by Loebis (1999), Oelim (2000), Sihotang et al. (2012), and Irwandi et al. (2019) reported that the water level of Lake Toba has gradually and significantly decreased over the last six decades. Furthermore, Irwandi et al. (2019) stated that there were specific periods of water level fluctuations marked by a significant decrease, which often occurred at scales from months to years (Acreman et al. 1993; Tanakamaru et al. 2004; Irwandi et al. 2019). The decrease and fluctuation of the Lake Toba water level, whose causes have not been explicitly stated by previous studies, have become intriguing topics among researchers due to their significant impact. According to Wesli (2017), the decline in Lake Toba's water level disrupts interisland transportation in the tourism sector. Meanwhile, in the industrial sector, its decrease significantly affects hydropower operations, which reduces the rate of aluminum production at PT Inalum (Acreman et al. 1993; Loebis 1999; Tanakamaru et al. 2004; Sihotang et al. 2012; Lukman 2017).

In a study on Lake Urmia in Iran, Jalili et al. (2016) stated that the El Niño Southern Oscillation (ENSO), climate change, and human activities were the main causes of the lake level's decline and fluctuation. Studies on a number of lakes around the world have shown that climate variability, such as ENSO, is the most significant phenomenon that causes an increase or decrease in lake levels, such as that observed in Lake Victoria in western Africa (Stager et al. 2007), the lakes of the Tibetan Plateau (Lei et al. 2019), Lake Volta in Ghana (Boadi and Owusu 2019), Lake Chad in central Africa (Okonkwo et al. 2014; 2015), and Lake Hawassa in Ethiopia (Belete et al. 2017). In addition to ENSO, the Atlantic Multidecadal Oscillation (AMO), Atlantic Meridional Mode (AMM), Indian Ocean Dipole (IOD), Pacific Decadal Oscillation (PDO), Pacific Quasidecadal Oscillation (QDO), Southern Oscillation Index (SOI), and North Atlantic Oscillation (NAO) have significant effects on the increase or decrease in the water levels of several lakes, such as the Great Lakes in the United States (Wang et al. 2010), Lake Urmia in Iran (Rezaei and Gurdak 2020), and Lake Chad in central Africa (Ndehedehe et al. 2016).

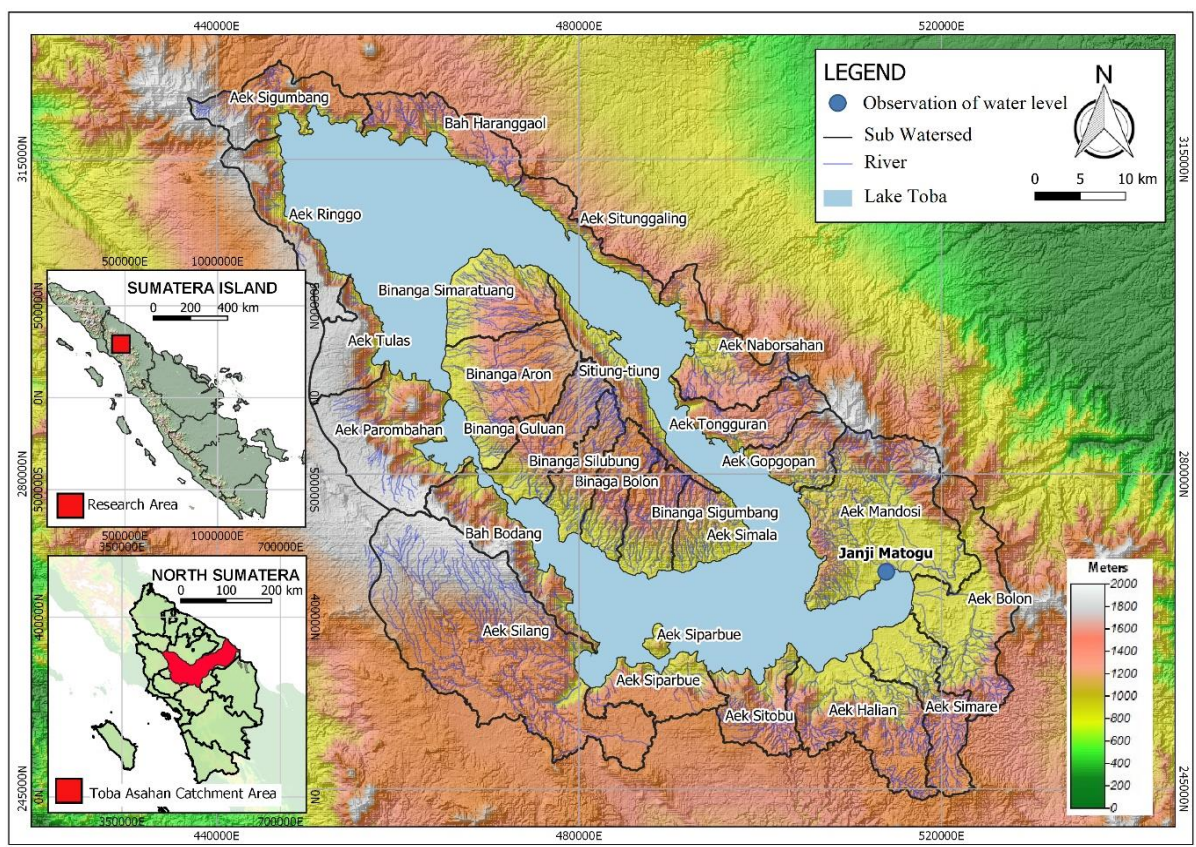

**Fig. 1:** Watershed area of Lake Toba is part of the Toba Asahan River basin in North Sumatra Province, Indonesia, within the Barisan Hills that stretch from the north to south on the northern part of Sumatra Island. Observation of the Lake Toba water level for this study was carried out at Janji Matogu, Toba Samosir Regency (indicated by the blue circle).

Several studies have stated that the increase in air temperature and decrease in rainfall significantly affect the decrease in lake levels. These phenomena were observed in Qinghai Lake in Qinghai Province, China (Zhang et al. 2011; Chang et al. 2017; Fang et al. 2019), lakes on the Tibetan Plateau (Huang et al. 2011; Song et al. 2014), Lake Michigan and Lake Huron in the United States (Argyilan and Forman 2003; Hanrahan et al. 2010), Lake Superior in the United States (Garcia and Townsend 2016), Lake Sibayi in southern Africa (Nsubuga et al. 2019), Lake Dianchi in China (Zhou et al. 2014), Lake Urmia in Iran (Arkian et al. 2018; Schulz et al. 2020; Dehghanipour et al. 2020a, b), Lake Issyk-Kul in Kyrgyzstan (Salamat and Abuduwaili 2015; Alifujiang et al. 2017), and Lake Dongping in Shandong Province, China (Rong et al. 2013). Furthermore, human activities are critical factors affecting decreases in lake levels. Examples of lakes whose water levels have been compromised by human activities are Lake Burdur in Turkey (Davraz et al. 2019), Lake Victoria in western Africa (Awange. 2008a; b; Okotto et al. 2018), Lake Urmia in Iran (Alizade et al. al. 2018; Alizadeh-Choobari et al. 2016; Khazaei et al. 2019), Lake Chad in central Africa (Mahmood et al. 2019), Lake Tana in Ethiopia (Minale 2019), and Lake Mead in the United States (Holdren and Turner 2010; Siyal et al. 2019).

In accordance with the studies on the decrease and fluctuation of lake levels in different parts of the world that are significantly related to ENSO, climate change, and human activities, the authors established the basis of this review on the decline and fluctuation of the water level of Lake Toba. This study aimed to critically examine the factors responsible for the decline in the Lake Toba's water level. The study also determined whether the instability of the lake was related to ENSO, climate change, or human interventions, thereby making it necessary to discuss lakes in other parts of the world that have been similarly affected. Finally, this study investigated existing prediction and projection models that simulate Lake Toba's water level to minimize the adverse impact and fluctuation through adaptation and mitigation efforts. The results are expected to mitigate existing problems and provide beneficial information for all stakeholders and future studies.

## 2. OBSERVATION OF THE LAKE LEVEL OF LAKE TOBA

The water volume and level of Lake Toba are influenced by the amount of rainfall that falls onto the lake's surface and inflow from surrounding rivers. According to the Indonesian government (2014), a total of 25 sub-watersheds flow into Lake Toba. Meanwhile, Acreman et al. (1993), Loebis (1999), and Oelim (2000) stated that Lake Toba experiences water loss due to evaporation, evapotranspiration, and outflow. Therefore, the level fluctuation is strongly influenced by the inflow and outflow of water, as shown in Fig. 2 (Loebis 1999).

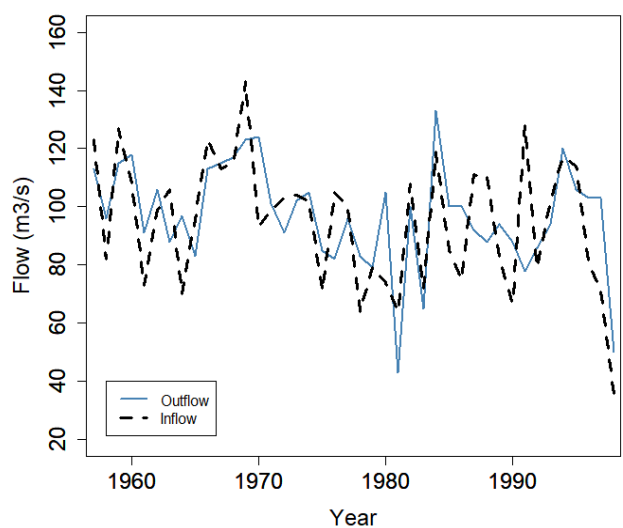

**Fig. 2:** Average annual inflow (dark brown dashed line) and outflow (light blue) of Lake Toba during the period of 1957-1998 (Loebis 1999).

Observations of Lake Toba's water level were carried out by PT Indonesia Aluminum in Janji Matogu, Toba Samosir District, as shown in Fig. 1. The first observation was conducted from 1914 to 1918, with a brief termination period due to the First World War. Observation resumed in 1920 and stopped in 1932 due to the Second World War. Finally, after Indonesian independence in 1945, the observation process resumed in 1957 and has been in existence since. Loebis (1999) stated that the unit used to determine the Lake's level in meters above sea level (masl).

Figure 3 shows a clear and significant downward trend in Lake Toba's water level fluctuations from 1957 to 2016. The study carried out by Irwandi et al. (2019) showed a decreasing trend in the water level of Lake Toba, with an average decrease of 2.4 cm per year over six decades. A significant decrease in the water level occurred from 1984 to 1987 (Acreman et al. 1993) and 1996 to 1997 (Tanakamaru et al. 2004), with decreases of 2.5 m and 2.0 m, respectively. In a recent study carried out by Irwandi et al. (2017a, b), using data obtained from 2015 to 2016, a water level decline of 2.0 m was observed in the lake. This result has raised great concern, because all existing studies have noted declines and fluctuations in Lake Toba's water level. However, these studies were unable to systematically explain and connect the existing indicators to explain the main cause and supporting factors associated with this phenomenon.

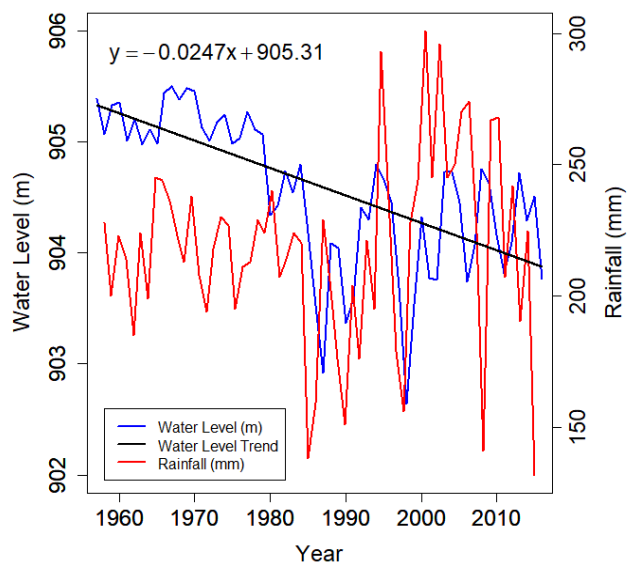

**Fig. 3:** Annual decline in Toba's lake level (blue line) and annual rainfall graph in the Lake Toba catchment area (red line) obtained using data from the Global Precipitation Climatology Center (GPCC) from 1957 to 2016. The solid black line represents the lake level trend, which has been decreasing at 2.4 cm per year for the last six decades (Irwandi et al. 2019).

## 3. DECREASE AND FLUCTUATION OF LAKE TOBA WATER LEVEL

Lakes are ecosystems that are most sensitive to variability and changes in climate (Catalan et al. 2013). Lei et al. (2019) stated that climate variability due to El Niño decreases rainfall and lowers the lake level. However, other studies reported different results (Belete et al. 2017). Climate change has generally caused the lake water level to decline (Khazaei et al. 2019), and natural lake ecosystems have responded to climate change by storing integrated information into sediments. These sediments allow paleolimnologists to document the climate changes experienced by the lake over thousands of years (Tranvik et al. 2009; Williamson et al. 2009). Human activities, such as irrigation practices and groundwater depletion, have led to the common disappearance of lakes around the world (Alizadeh-Choobari et al. 2016). According to Torabi and Kløve (2015), these changes in lake levels are influenced not only by climate but also by intensive land use, thereby directly affecting hydrological, ecological, and ecosystem support. In the case of Lake Toba, the factors in the following subsections have been indicated as the factors responsible for the decrease and fluctuation in its water level.

## 3.1 ENSO

Climate variability is closely related to ENSO activity in the Pacific Ocean (Naylor et al. 2002). El Niño is a phenomenon that occurs in the Pacific Ocean. This phenomenon occurs if the anomaly in sea surface temperature (SST) during 5 months in Niño region 3.4 (5° N-5° S, 120°–170° W) exceeds 0.4°C for 6 months or more (Trenberth, 1997). In Indonesia, the June-July-August (JJA) and September-October-November (SON) periods are indicated by a significantly negative rainfall anomaly. This anomaly is significant only in southern and eastern Indonesia in the December-January-February (DJF) period. Meanwhile, in the March-April-May (MAM) period, the effect is generally insignificant, except in the northern Sumatra and Sulawesi regions (Supari et al. 2018).

ENSO is a global phenomenon that has played an essential role in the reduction of rainfall in the Lake Toba catchment area (Loebis 1999; Irwandi et al. 2017a, 2018, 2019), resulting in a decrease in the water level of Lake Toba. During El Niño in 1982/1983, 1986/1987, 1991/1992, 1997/1998, and 2015/2016, the dry season in the Lake Toba watershed area was more arid and lasted longer, i.e., for 3 to 5 decades, in which the rainfall reduction ranged from 50 to 100 mm in the DJF, JJA, and SON periods. In contrast, rainfall was reduced by 0-100 mm in the MAM period, with a 0-50 mm increase in the western region of the Lake Toba catchment area, as shown in Fig. 4 (Irwandi et al. 2018). This pattern was proven by the results of a recent study carried out by Irwandi et al. (2019), which stated that the rainfall anomaly in the Lake Toba catchment area was strongly correlated with the active El Niño period. Rainfall data indicated monthly decreases from 0 to 150 mm in 1958, 1963, 1968, 1977, 1991, 1992, 1995, 1997/98, and 2015. The study also observed a significant decrease in rainfall between

1997 and 2015. From the data presented above, it was concluded that there was a relationship between the El Niño phenomenon and climate variability and the decrease in the Lake Toba water level. However, more detailed studies are needed to firmly prove whether El Niño-related climate occurrences are the only cause of this phenomenon in Lake Toba.

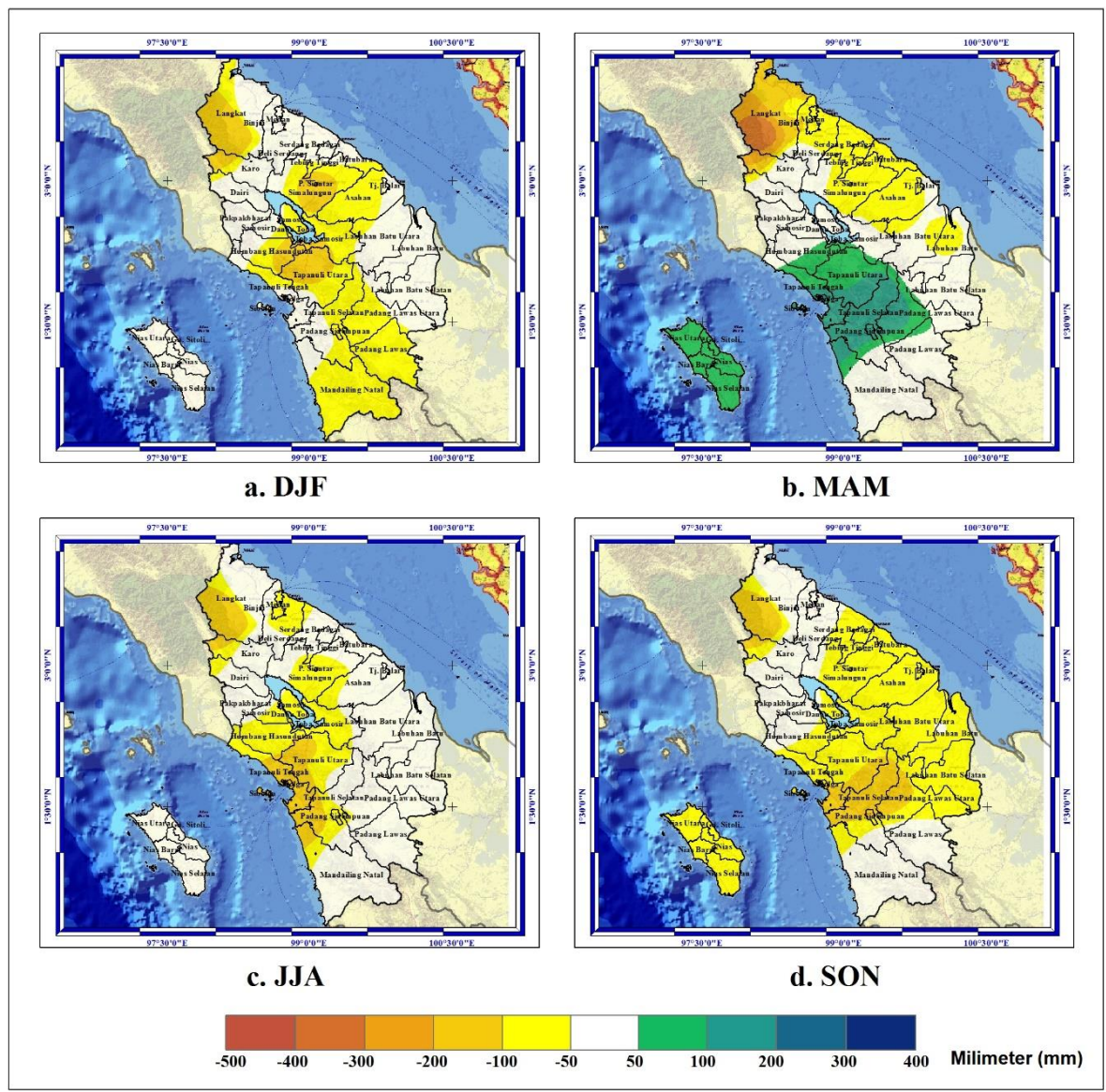

**Fig. 4:** Spatial patterns of rainfall anomalies during El Niño occurrences in 1983/1984, 1986/1987, 1991/1992, 1997/1998, and 2015/2016 in North Sumatra Province, Indonesia, in a DJF, b MAM, c JJA, and d SON periods. This figure shows a decrease in seasonal rainfall in North Sumatra, ranging from 50 to 100 mm, except on the western coast of North Sumatra during the MAM period, which had increased rainfall ranging from 0 to 150 mm (Irwandi et al. 2018).

Previous studies of several other lakes have shown that rainfall is generally the leading cause of lake level fluctuations (Hwang et al. 2005; Kebede et al. 2006; Zhang 2011; Soja et al. 2013; Muvundja et al. 2014; Song et al. 2014; Yujie Yuan et al. 2015; Okonkwo et al. 2015; Yao et al. 2018; Vanderkelen et al. 2018). Awange et al. (2008a) proved that 80% of the inflow of Lake Victoria in western Africa was derived from rainfall; therefore, the contribution of rainfall is crucial and cannot be ignored. A similar conclusion was stated in the study carried out by Cui and Li (2016), which indicated that rainfall and evaporation were the main factors affecting the Qinghai Lake levels in China.

The study carried out by Stager et al. (2007) and Williams et al. (2015) stated that Lake Victoria in western Africa has also been affected by tropical rainfall and enhanced by the interaction of SST and atmospheric circulation systems in the Pacific and Indian Oceans, including ENSO. Seasons with lake level deficiencies occurred in the periods of JJA (1967), SON (1972), SON (1980), DJF (1986), SON (1995), and JJA (1997), in which the seasonal periods were identified as active El Niño years except for SON (1980) (Awange et al. 2008a). Lei et al. (2019) stated that the Tibetan Plateau's lakes shrunk significantly due to reduced rainfall during the 2015/2016 El Niño. Lake Volta in Ghana also exhibited this precipitation variability, with ENSO contributing to interannual fluctuations from 1970 to 2010 (Boadi and Owusu 2019). Meanwhile, Okonkwo et al. (2014) reported that Lake Chad in central Africa showed negative relationships between ENSO and water level, river discharge, and rainfall. This result was corroborated by the study carried out by Ndehedehe et al. (2016), in which ENSO, AMO, and AMM were correlated with the extreme rainfall conditions in the area. A similar condition occurred in Lake Hawassa in Africa, in which the water level increased and decreased during El Niño and La Niña, respectively. Furthermore, a strong El Niño increased the lake level in 1997/98, while a decrease occurred in 1975 due to a strong La Niña occurrence (Belete et al. 2017). However, the research carried out by Yin et al. (2009) found no significant relationship between ENSO and catastrophic flooding or drought in Taihu Lake in China.

In addition to ENSO, variability has shown a significant connection with lake level. The variabilities of NP and PDO were observed as significant factors affecting the Lake Qinghai water level changes in China. The lake level variations of the Great Lakes in the United States were also closely related to the SST anomaly in the tropical central Pacific, also known as QDO (Wang et al. 2010). Similarly, the changes in Urmia's lake level in Iran exhibited a significant relationship with activities in the Pacific Ocean (i.e., ENSO and PDO) (Jalili et al. 2016; Rezaei and Gurdak 2020).

## 3.2 Climate change

Changes in lake characteristics act as climate change indicators (Adrian et al. 2009; Tranvik et al. 2009; Williamson et al. 2009). Climate change refers to an alteration in climatic conditions over an average or prolonged period, usually decades or more, due to an increase in $CO_2$ gas, also known as the greenhouse effect (IPCC 2013). Furthermore, water resources worldwide have faced formidable challenges in recent decades characterized by a decrease and increase in rainfall and temperature, respectively. Most regions globally anticipate the adverse impacts of climate change, especially on water resources (IPCC 2013). This process is also related to lake ecosystem sustainability effects, in which increasing air temperature results in increased lake evaporation (Awange et al. 2008a).

Rainfall and temperature are the most common indicators that affect lake levels, because they provide more tangible contributions, and we have excellent and detailed knowledge regarding their impacts on specific scales (Reyes-García et al. 2016). Research carried out on the analysis of climate change by Irwandi et al. (2017a, b) in the Lake Toba catchment area showed an increase in temperature and a decrease in rainfall from 1997 to 2016. These data were further concluded by the Parapat Geophysical Station after analyzing temperature trends and rainfall in the Lake Toba catchment area. In January, the average, maximum, and minimum temperatures exhibited annual increasing trends of 0.02°C, 0.05°C, and 0.004°C per year, respectively, as shown in Fig. 5a. Meanwhile, the rainfall during the dry (January-June) and rainy (July-December) seasons showed decreasing annual trends of 3.18 mm and 16.33 mm, respectively, as shown in Fig. 5b. This result proves that climate change has indeed caused variations in the Lake Toba watershed area. However, further studies need to be conducted, particularly in relation to the decrease in the lake's water level.

In reviewing the studies mentioned above, the changes in climate that occurred in the Lake Toba area during the observation period were defined as a significant increase in air temperature and a decrease in rainfall. This condition also occurred on the Tibetan Plateau (Huang et al. 2011), in Lake Superior in the United States (Garcia and Townsend 2016), in Lake Sibayi in Africa (Nsubuga et al. 2019), in Lake Dianchi in China (Zhou et al. 2014), in Lake Urmia in Iran (Alizadeh et al. 2016; Nourani et al. 2019; Schulz et al. 2020; Dehghanipour et al. 2020a, b), and in Lake Poyang in China (Sun et al. 2013). Furthermore, global climate change impacts have been observed to affect the water levels of lakes in all countries. For example, Lake Urmia in Iran has experienced a dramatic decline in water levels of approximately 8 m since 1995 (Khazaei et al. 2019), and the Qinghai Lake level in China annually decreased by 7 cm from 1956 to 2009 while experiencing an annual increase in air temperature of 0.03°C (Zhang 2011; Fang et al. 2019). Over the past 80 years, the water level fluctuation of Lake Issyk-Kul in Kyrgyzstan occurred due to an increase in temperature from 0.7 to 1.5°C and a decrease in rainfall of 45-130 mm (Salamat and Abuduwaili 2015). In Lake Urmia in Iran, the average high rainfall decreased by 9.2%, with an increase in the average maximum temperature of 0.8°C over the past four decades (Delju et al. 2013). Seasonal change, which occurred during winter and spring, in Dongping Lake in China became more arid from 2003 to 2010 due to the annual evaporation increase of 4.55 mm/year (Zhang 2011).

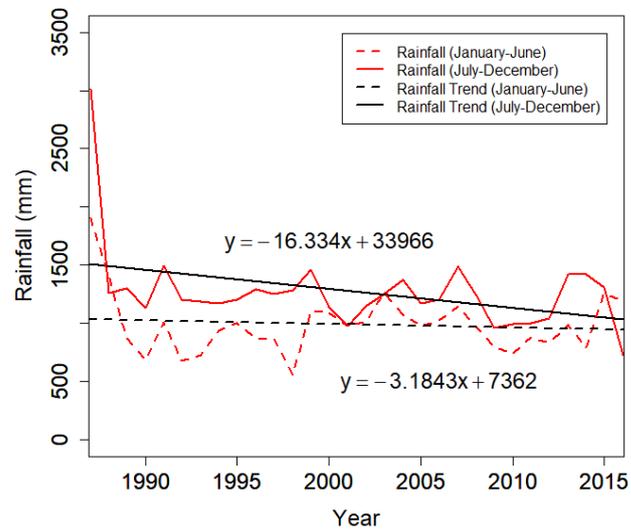

(a)

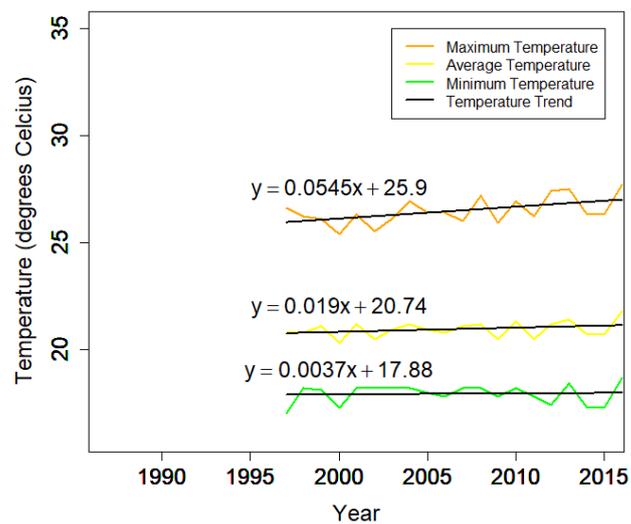

(b)

**Fig. 5** a Rainfall trend for the January-June (red dotted line) and July-December (red line) periods. The rainfall during the dry season (January-June) shows a decreasing annual trend of 3.18 mm, and the rainy season (July-December) exhibits a decreasing annual trend of 16.33 mm in the Lake Toba catchment area. b Trends in maximum (light brown line), average (yellow line), and minimum (green line) temperatures. The average temperature exhibits an increasing trend of 0.02°C per year, the maximum temperature shows an increasing annual trend of 0.05°C, and the minimum temperature shows an increasing annual trend of 0.004°C (Irwandi et al. 2017a, b).

## 3.3 Human activities

The population surrounding Lake Toba has continually increased, which tends to increase the need for freshwater supply and lake water use (Wesli 2017; Aziz et al. 2020). A recent study carried out by Aziz et al. (2020) stated that land clearance for new households increased significantly from 1990 to 2018, with the most significant rate observed in the Toba Samosir Districts located adjacent to Lake Toba. From 1990 to 1997, the open land area was only 7.54 $km^2$, while in 2013-2018, it increased to 68.10 $km^2$. Lake Toba's largest surface area (1133.88 $km^2$) was found from 1990 to 1997, while the smallest surface area (1119.20 $km^2$) was observed from 1998 to 2004. Figure 6 shows a strong correlation between the open land and Lake Toba surface areas with a value of -0.77, which indicates that the increase in open land area creates a decrease in lake surface area.

Industrial activities, such as an excessive use of water for hydroelectric power (Acreman et al. 1993) and deforestation (Aswandi and Kholibrina 2017; Saragih and Sunito 2001; Sianturi 2004; Sihotang et al. 2012), have also contributed to the decrease in Lake Toba's water level. However, no prior studies have been devoted to investigating their effects on lake level declines and fluctuations.

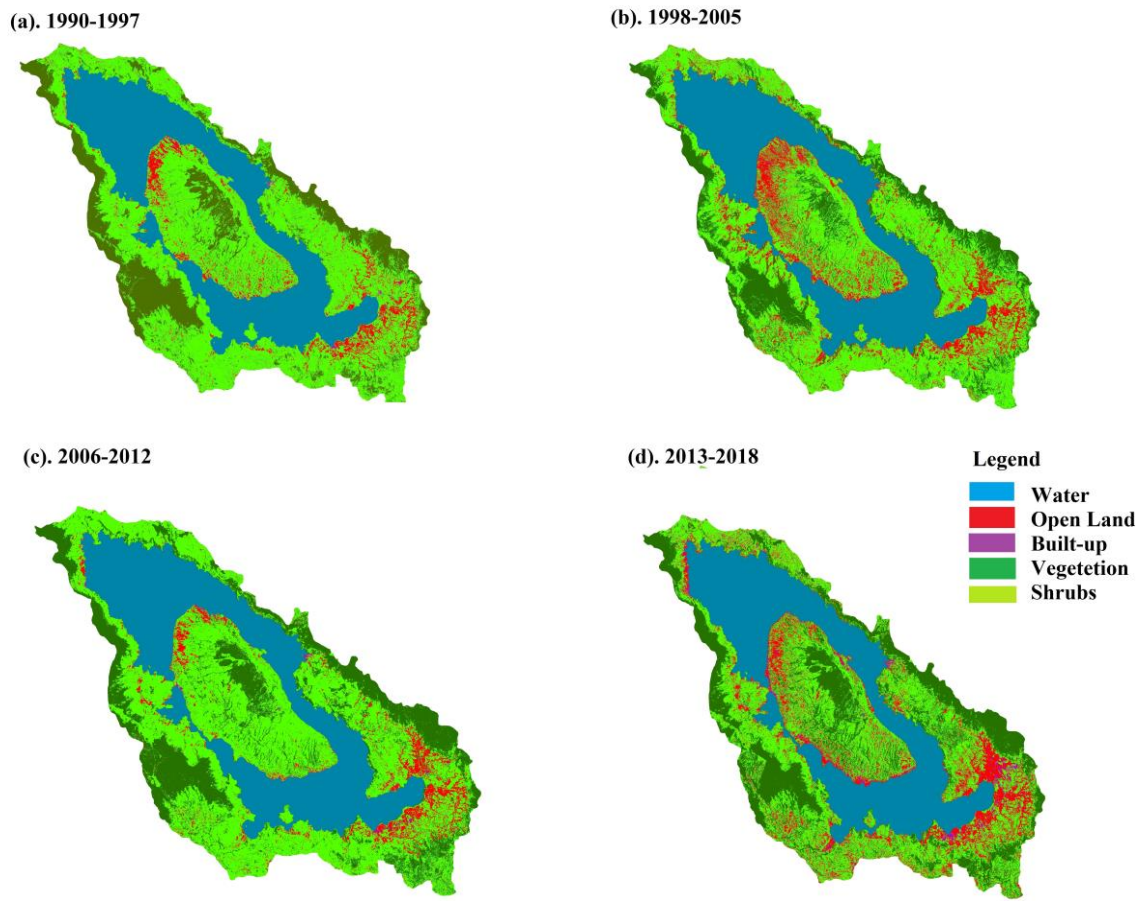

**(a). 1990-1997**

**(b). 1998-2005**

**(c). 2006-2012**

**(d). 2013-2018**

**Legend**
- **Water**
- **Open Land**
- **Built-up**
- **Vegetetion**
- **Shrubs**

**Fig. 6:** Changes in the open land area during the 1990-2018 period in the Lake Toba catchment area (Aziz et al. 2020).

Studies carried out by Bashirian (2020), Fan (2020), and Sinyukovich and Chernyshov (2019) focused on human activities associated with dam regulation and their use for agricultural irrigation and hydropower generation. Dehghanipour et al. (2019, 2020a, b) developed a model to determine the potential water management strategies sourced from lakes for agricultural land irrigation including Lake Urmia in Iran. In this research, the water used was based on the balance in anticipation of the lake level declining as a result of human activities related to agricultural activities. Furthermore, Pokhrel et al. (2017) carried out research illustrating the important roles played by humans in maintaining water resource sustainability for future usage. The impacts of human activities contributed to a decrease in lake water level. For instance, the surface area of Lake Sibayi in Africa was reported to have shrunk by 20% from 1992 to 2016 (Nsubuga et al. 2019), and Lake Burdur in Turkey dramatically decreased from 210 km$^2$ to 131 km$^2$ between 1975 and 2016, a decrease of 37% (Davraz et al. 2019). Furthermore, Lake Victoria in Africa decreased by 31% from 1984 to 2010 (Okotto et al. 2018), while the surface area of Lake Urmia in Iran shrank by 42.2% in 2011 compared to 1998, with an annual decrease of 44.3 cm from 1995 to 2009. According to Alizadeh-Choobari et al. (2016) and Ghale et al. (2018), anthropogenic and climatic factors contribute approximately 80% and 20% of the impacts on Lake Urmia, respectively.

Other studies have found that climatic factors are not the only cause of the dramatic decline in the water level of Lake Victoria in western Africa. Awange et al. (2008b) stated that the leading cause is the increase in the Owen Falls Dam volume. This finding was confirmed by the studies carried out by Bashirian et al. (2020) and Fan et al. (2020), who stated that regulations of the dam have indirectly created a significant decrease in the lake level. The surface elevation of Lake Mead has also decreased by 40 m since 1999 in accordance with prolonged drought and an increase in water supply demand due to population growth (Holdren and Turner, 2010; Siyal et al. 2019). In the case of Lake Chad in Africa, 66% of the total decrease in river flow was due to human activities, while 34% was due to climate variability (Mahmood et al. 2019). The water level of Lake Tana in Ethiopia has also shown an extreme and rapid decline in recent years due to the change in land use and non-optimal water regulation related to the Chara Dam (Minale 2019). Sinyukovich and Chernyshov (2019) stated that the decrease in Lake Baikal's water level in Russia was due to the discharge of hydroelectric power plant activity into the lake.

## 4. WATER LEVEL SIMULATION OF LAKE TOBA AND UTILIZATION OF GENERAL CIRCULATION MODELS (GCMs)

Lake level simulation is an approach used to predict a future water supply. Although there is relatively large uncertainty in simulation-based estimation, this analysis needs to be conducted as an input for future water management. Sihotang et al. (2012) carried out simulation research of Lake Toba's water balance from 2017 to 2057. The parameters used in the simulation included population growth, water runoff, soil water absorption, land-use control, and water discharge regulation on the Asahan River. The result indicated that Lake Toba's future water inflow would be significantly higher than its outflow, with the desired lake level at 905.30-905.50 masl and a difference between its highest and lowest elevations of 0.60 m.

This result is in line with the research carried out by Wesli (2017) in relation to North Sumatra integrated water management, which uses Lake Toba as a freshwater supply. The simulation was conducted until 2032 using parameters of increasing population and mainstay discharge. The results showed that the required output discharge was 82.37 m$^3$/s less than the inflow rate of 209.09 m$^3$/s. A surplus supply of approximately 121.79 m$^3$/s and 87.3 m$^3$/s came from rainfall and the surrounding rivers, respectively. Therefore, from these two studies, it was concluded that the water level in Lake Toba water could be optimized without affecting its future. However, further studies need to be carried out using different methods and more parameters.

High-resolution general circulation models (GCMs) are the most commonly used current model to describe future climate projections under various scenarios (Akurut et al. 2014). GCMs have also been used to simulate future water balances and lake levels in several studies, such as the research carried out by Soja et al. (2013) on a water balance simulation in Lake Neusiedl. Soja's study found that from 2035 to 2065, the water level is likely to be lower than the current level. A study carried out by Mahsafar et al. (2014) stated that Lake Urmia's water level in Iran was projected to decline by 4.60 m from 2010 to 2100. Jamali and Rhazi (2018) stated that Lake Aguelmam Sidi Ali's water level in Morocco would experience a continuous decline until 2044. This research explains why climate projections can explain the critical conditions of a decrease in future lake levels. High-resolution projections of simulated lake levels are expected to incorporate the main controlling factors of the lake levels in addition to the emission scenario (Vanderkelen et al., 2018).

Furthermore, studies by Yagbasan et al. (2012) and (2017) indicated that the water levels of Lake Mogan and Lake Eymir in Turkey are likely to undergo significant decreases and experience potential drought based on various climate projection scenarios. The simulation showed seasonal air temperature increases of 6°C and 4°C for scenarios A2 and B1, respectively, by 2100. Autumn and spring are likely to experience increases, while the lowest trend of increasing air temperature will occur in winter, as shown by Yagbasan et al. (2012). In scenario A2, the rainfall in winter and autumn is illustrated by a downward trend, while the spring and summer rainfall decreases significantly. In scenario B1, seasonal rainfall exhibits a uniform downward trend, except in spring, in which a relatively more significant decline is expected. According to Yagbasan et al. (2012), the decrease in the rainfall projected in scenario B1 is significantly lower than that projected in scenario A2. The results of scenarios A2 and B1 indicate that long-term changes in rainfall and temperature tend to significantly decrease the water level, thereby causing the lake to become drier during the dry season.

Several climate projection scenarios that show a decrease in water levels and drought potential have been carried out on Lake Urmia in Iran by Mahsafar et al. (2014) and Sanikhani et al. (2018) as well as on the Great Lakes in the United States by Angel and Kunkel (2010). Meanwhile, there have been other climate projection scenarios that show increased air temperature, e.g., Lake Guiers in Senegal by Tall et al. (2017), the Great Lakes in the United States by Rao et al. (2012), and Lake Chad in Africa by Mahmood et al. (2019), with increases ranging from 10 to 25% for rainfall or 0.5-2.0°C for air temperature. A combination of both increased rainfall and air temperature represents a different phenomenon due to the presence of another projection (Lee et al. 2012) in the Yongdam Reservoir and Lake Victoria. These results indicated that different climate projections were expected for each lake.

Previous studies agreed that predicted future climatic conditions pose a threat to water availability based on the above reviews. A similar result was found by Dai (2011), whose climate projection scenarios showed an increase in drought in most of Africa, southern Europe, the Middle East, America, Australia, and Southeast Asia in the twenty-first century. However, Sihotang et al. (2012) and Wesli (2017) carried out studies based on the simulation of Lake Toba's water balance and showed a potential surplus until the mid- twenty-first century. This research is important for determining the related methods, data series, and parameters used as the research indicators to arrive at conclusions based on global simulations and published climate scenarios. Prudent anticipation of these changes in the lake management system is needed for future mitigation efforts. Furthermore, by considering the climate projections capable of influencing the lake levels, it is expected that relevant stakeholders, together with the community, can maintain and preserve the ecosystem in the Lake Toba catchment area by anticipating adverse impacts likely to occur in the future.

## 5. CONCLUSION

This study conducted a systematic and critical review of published scientific reports on the Lake Toba water level and compared them with similar phenomena to other lakes around the world. Therefore, based on these reviews, the following conclusions were obtained.

Over the last six decades, Lake Toba's water level has decreased significantly by 2.4 cm /year from 1957 to 2016. Studies have shown that global climate dynamics due to variables in the Pacific Ocean, such as ENSO, reduced the amount of rainfall in the Lake Toba watershed area, thereby causing the decrease in lake level fluctuations within the last four decades.

The increasing trend in the maximum and average temperatures and the decrease in rainfall in the Lake Toba area are due to climate change. Other factors essential in the decline and fluctuation of the Lake Toba water level are industrial and household activities, such as water discharge from hydropower plants, environmental damage, and a significant increase in the population. Therefore, to confirm these causes, a more comprehensive and in-depth study is needed.

A simulation study of Lake Toba's water balance was carried out with a projection period until the mid-21st century. The study concluded that there was a surplus condition, with predicted input and output discharges of 209.9 m$^3$/s and 82.37 m$^3$/s, respectively. This result indicates that Lake Toba's water level is within the normal limits at 905.30-905.50 m. Therefore, an improvement in water supply management is necessary to maximize its usage.

Climate variability and changes in human activities have also been found to be the factors responsible for fluctuations or decreases in lake levels by various global studies. These factors need further comprehensive and systematic studies to firmly determine the most relevant factor in the specific case of Lake Toba.

Climate projections have yielded several scenarios that predict the possibility of extreme climate anomalies. This constitutes an additional challenge, and therefore, it is a top priority to obtain climate projection information on a spatial scale with a detailed resolution. The Indonesian government and all stakeholders can use this information to carry out mitigation, adaptation, and anticipation activities against possible future drought.

**ABBREVIATIONS**

ENSO: El Niño Southern Oscillation; AMO: Atlantic Multi-decadal Oscillation; AAM: Atlantic Meridional Mode; IOD: Indian Ocean Dipole; PDO: Pacific Decadal Oscillation; QDO: Pacific Quasi-decadal Oscillation; SOI: Southern Oscillation Index; NAO: North

Atlantic Oscillation; DJF: December, January, February; MAM: March, April, May; JJA: June, July, August; SON: September, October, November; IPCC: The Intergovernmental Panel on Climate Change; GCMs: General Circulation Models.

## AVAILABILITY OF DATA AND MATERIALS

The data and materials reported in this review are from the published literature.

## COMPETING INTERESTS

The authors declared no conflict of interests.

## FUNDING

This research was funded by the Universitas Indonesia through the PUTI Doktor 2020 Grant program (BA-801/UN.2.RST/PPM.00.03.01/2020) and PUTI Q1 2020 Grant (BA-1080/UN2.RST/PPM.00.03.01/2020).

## AUTHORS' CONTRIBUTIONS

TM, MSR designed the research, while HI reviewed articles and drafted the manuscript. All authors read and approved the final manuscript.

## ACKNOWLEDGMENTS

The authors are grateful to the Indonesian Center for Education and Training of the Meteorology, Climatology, and Geophysics Agency for providing scholarships for the Doctoral program. The University of Indonesia also funded this research through the 2020 PUTI Doctoral Grant (BA-801/UN.2.RST/PPM.00.03.01/2020) and PUTI Q1 2020 grant (BA-1080/UN2.RST/PPM.00.03.01/2020).